\documentstyle[11pt,moriond,epsfig]{article}

\def\Journal#1#2#3#4{{#1} {\bf #2}, #3 (#4)}

\def\NPB{{\em Nucl. Phys.} B}
\def\PLB{{\em Phys. Lett.}  B}

\def\PRD{{\em Phys. Rev.} D}

\def\CPC{\em Comp. Phys. Comm.}
\def\EPJ{{\em Euro. Phys. J.} C}

\def\be{\begin{equation}}
\def\ee{\end{equation}}
\def\bea{\begin{eqnarray}}
\def\eea{\end{eqnarray}}

\begin{document}
\vspace*{4cm}
\title{$\mathbf W$/$\mathbf Z$ + $\mathbf B{\overline{B}}$/JETS
 AT NLO USING THE MONTE CARLO MCFM}

\author{ J.M. CAMPBELL }

\address{Theory Department, Fermilab, PO Box 500,\\
Batavia, IL 60510, USA}

\maketitle\abstracts{
We summarize recent progress in next-to-leading order QCD calculations
made using the Monte Carlo {\tt MCFM}. In particular, we focus
on the calculations of $p{\bar p} \to Wb{\bar b}$,~$Zb{\bar b}$
and highlight the significant corrections to background
estimates for Higgs searches in the channels $WH$ and $ZH$ at the
Tevatron. We also report on the current progress of, and strategies
for, the calculation of the process $p{\bar p} \to W/Z + 2$~jets.}

\section{{\tt MCFM} Background}
 
With the advent of Run II at the Tevatron we will be able to
perform detailed studies of femtobarn-level processes for the
first time. The production of final states involving heavy
quarks, leptons and missing energy is particularly interesting
since these are common signatures for new physics beyond the
Standard Model. For example, a light Higgs in the mass range
$110$~GeV$ < m_H < 140$~GeV predominantly decays into a
$b{\bar b}$ pair. In order to assess search strategies and
perform meaningful comparisons with the data, a solid knowledge
of the Standard Model backgrounds is necessary. In an attempt 
to fill this need, the Monte Carlo program {\tt MCFM}
aims to provide a unified description of many of these
femtobarn-level processes at next-to-leading order (in the
strong coupling, $\alpha_S$) accuracy. In many cases, the
extension to NLO is made feasible by the recent calculations
of virtual matrix elements involving a vector boson and four
partons.~\cite{4jetvirt}

The philosophy behind {\tt MCFM} is somewhat similar to that
of a general-purpose showering Monte Carlo such as
Pythia~\cite{Pythia} (although here we are of course working at
a fixed order in ${\alpha_S}$). One specifies a set of basic
parameters and cuts, selects a process number from a table of
included processes and then the Monte Carlo {\tt MCFM}
produces a set of relevant distributions.
The main processes currently implemented at next-to-leading
order are shown in Table~\ref{table:processes}, with various leptonic
and hadronic decays of the bosons included as further
sub-processes.
Version 1.0 of the program is available for download at
{\tt http://www-theory.fnal.gov/people/campbell/mcfm.html}.
\begin{table}[ht]
\caption{The main processes included in {\tt MCFM} at next-to-leading
order.}
\begin{center}
\begin{tabular}{|ll|}
\hline
$p {\bar p} \to W^\pm/Z$ & \qquad  $p {\bar p} \to W^+ + W^-$ \\
$p {\bar p} \to W^\pm + Z$ & \qquad $p {\bar p} \to Z + Z$\\
$p {\bar p} \to W^\pm/Z + H$ & \qquad $p {\bar p} \to W^\pm/Z + \mbox{1
jet}$ \\
$p {\bar p} \to W^\pm + g^\star \, ( \to b {\bar b} )$ & \qquad
$p {\bar p} \to Z b {\bar b}$ \\
\hline
\end{tabular}
\end{center}
\label{table:processes}
\end{table}

\section{Higgs Search Using {\tt MCFM}}
 
Studies using lowest-order Monte Carlos and other event generators
show that for a Higgs in the mass range of $100$-$130$~GeV,
the most promising channels for discovery at Run II are
associated Higgs production,~\cite{stangemw}
\begin{eqnarray*}
&& p{\bar p} \longrightarrow W (\to e \nu) {H (\to b{\bar b})}, \\
&& p{\bar p} \longrightarrow Z (\to \nu {\bar \nu}, \ell {\bar \ell}) {H
(\to b{\bar b})}.
\end{eqnarray*}
This mass region is particularly interesting in the light of recent
hints from LEP2 suggesting a Higgs mass $m_H = 115$~GeV.

Considering first the $WH$ signal, the main backgrounds are,

\vspace*{-0.2cm}
\begin{minipage}{3in}
\begin{eqnarray*}
p{\bar p} & \longrightarrow & W \, {g^\star (\to b{\bar b})} \\
p{\bar p} & \longrightarrow & W \, {Z/\gamma^\star (\to b{\bar b})}
\end{eqnarray*}
\end{minipage}
\hspace*{-0.2cm}
\begin{minipage}{1.5in}
\begin{eqnarray*}
p{\bar p} & \longrightarrow & t(\rightarrow bW^+) \bar{t}(\rightarrow
\bar{b}W^-)\\
p{\bar p} & \longrightarrow & W^{\pm *} (t(\rightarrow bW^+) \bar{b}) \\
qg & \longrightarrow & q^{\prime} t(\rightarrow bW^+) {\bar b}.
\end{eqnarray*}
\end{minipage}
\\
\vspace*{0.2cm}

\begin{figure}
\begin{center}
\hspace*{1cm}\epsfig{file=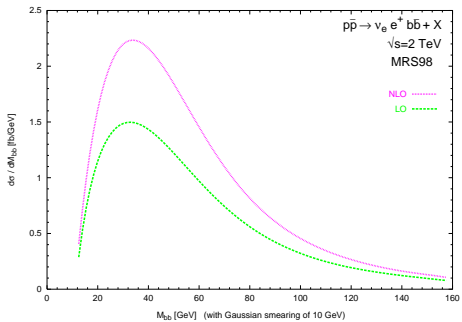,width=6cm}
\end{center}
\caption{The $m_{b{\bar b}}$ distribution of the
$W^+(\to \nu_e e^+) \, b{\bar b}$
background to the Higgs signal, at LO and NLO.}
\label{fig:wbbmbb}
\vspace*{-0.1cm}
\end{figure}

The signal, $W g^\star$ and $W Z$ backgrounds are calculable at
the one-loop level in {\tt MCFM} and the remainder at lowest order.
In order to study the largest background - the continuum $W g^\star$ -
we introduce a set of standard cuts from the
literature,~\cite{higgswg} use a double $b$-tagging efficiency of
$\epsilon_{b\bar{b}}=0.45$ and employ the MRS98 set of
parton distribution functions.~\cite{mrs98}
When searching for the Higgs
signal, we are interested in the cross-section as a function of
the $b{\bar b}$ pair mass, $m_{b{\bar b}}$. This distribution
is compared at lowest order and next-to-leading order in
Figure~\ref{fig:wbbmbb}, with a renormalization and factorization
scale $\mu = 100$~GeV.
From examining this figure, one can see that the effect of the
next-to-leading corrections is to increase the distribution
by a factor of approximately $1.5$ throughout, thus changing
the shape very little.

\begin{figure}
\begin{center}
\epsfig{file=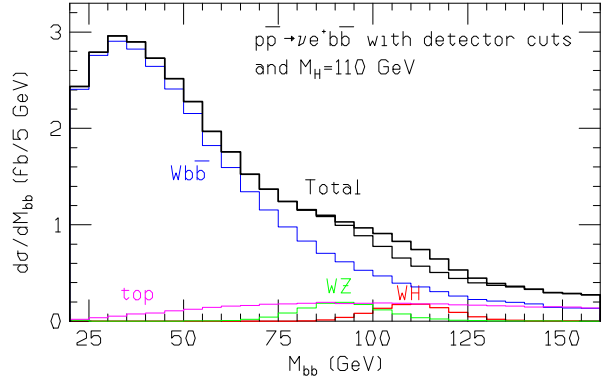,width=5cm,angle=90}
\end{center}
\caption{Parton-level signal and backgrounds assuming $m_H=110$~GeV.
The uppermost curve represents the signal+backgrounds, with the
(almost identical) lower curve representing just the sum of all the
backgrounds.}
\label{fig:whnlo}
\vspace*{-1.2cm}
\end{figure}

The results of a parton-level study for $m_H=110$~GeV,
with all but the small $t{\bar t}$ and single-top backgrounds
calculated at next-to-leading order, are summarized in
Figure~\ref{fig:whnlo}.
It is clear from the figure that the extraction of the Higgs signal
requires detailed knowledge of the normalization and the
kinematics of the backgrounds.

\begin{figure}
\begin{center}
\hspace*{-3cm}\epsfig{file=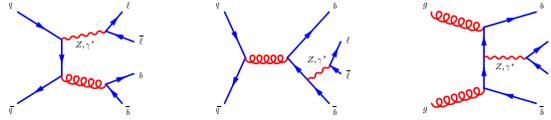,width=7.5cm}
\end{center}
\caption{Sample diagrams contributing to the process
$p{\bar p} \to Zb{\bar b}$. Diagram (c) represents a
gauge-invariant set that does not appear at all in the
$Wb{\bar b}$ calculation.}
\label{fig:zbbdiags}
\vspace*{-0.2cm}
\end{figure}

\begin{figure}
\begin{center}
\epsfig{file=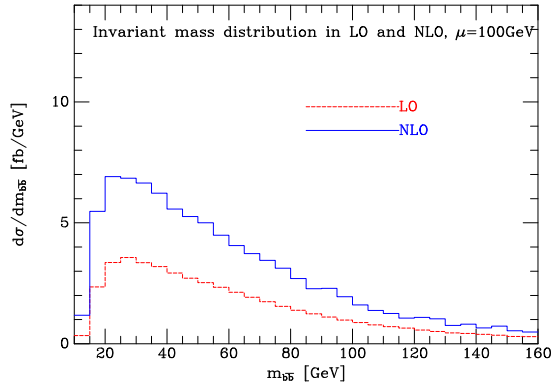,width=5cm,angle=-90}
\end{center}
\caption{The $m_{b{\bar b}}$ distribution for
the final state $Z (\to 3 \times \nu {\bar \nu}) \, b {\bar b}$
at leading and next-to-leading order.}
\label{fig:zbbnlo}
\vspace*{-0.2cm}
\end{figure}

It is straightforward to perform a similar search in the
$ZH$ channel.~\cite{jmcrkezbb} Here we shall concentrate on the
$Z (\to \nu {\bar \nu}) b{\bar b}$ background, calculated at
next-to-leading order with
{\tt MCFM}, and compare with the $Wb{\bar b}$ calculation above.
The Feynman diagrams required for this calculation - shown
in Figure~\ref{fig:zbbdiags} - are very similar to
the previous case, with additional
contributions from $gg$ initial states.
These $gg$ initial states are important for the Higgs search since
they can readily produce a $b{\bar b}$ pair with a large invariant mass.
For a conventional scale of $100$~GeV, the $m_{b{\bar b}}$
distribution is shown in Figure~\ref{fig:zbbnlo}.
The effect of the radiative corrections is to produce
a large $K$-factor $\approx 1.8$
in the region of interest, $m_{b{\bar b}}$ from $100$-$130$~GeV.

\section{$W + 2$~jets: Work in Progress}
 
The $p{\bar p} \to W + 2$~jets process can be viewed
as an extension of the $Wb{\bar b}$ and $Zb{\bar b}$ calculations
that have already been discussed in the previous sections.
There are of course extra parton configurations that we must count,
but the basic matrix elements (modulo couplings) are the same.
Specifically, the $Wb{\bar b}$ calculation contains all the
diagrams relevant for $q{\bar q} \to W + q^\prime{\bar q^\prime}$;
the $Zb{\bar b}$ process, together with crossings,
provides the $gg \to W + q{\bar q}$, $q{\bar q} \to W + gg$,
$gq \to W + gq^\prime$
and similar sub-processes. A further complication is that the
contribution from the diagrams that include real
radiation must incorporate the extra singularities due to more
instances of soft or collinear gluons and collinear quark pairs.
In order to simplify the calculation - and to reduce the required
computation time - we have decided to employ a
colour-decomposition of the matrix
elements. In this procedure the matrix elements are expressed as 
an expansion in $1/N$ (where $N$ is the number of colours, $3$),
with the hope that the sub-leading terms
are small compared to the leading term. Performing the colour
expansion we obtain ($V = W^\pm$, $Z$),
\begin{eqnarray*}
|{\cal M}_{NLO} (Vq{\bar q}gg)|^2 & \sim &
 {1} \times {\cal G}_0
 +\frac{1}{N^2} \times {\cal G}_2
 +\frac{1}{N^4} \times {\cal G}_4 \\
|{\cal M}_{NLO} (Vq{\bar q}Q{\bar Q})|^2 & \sim &
 {\frac{1}{N}} \times {\cal Q}_1
 {+\frac{1}{N^3}} \times {\cal Q}_3
 + \delta_{qQ} \,  \left( 1 \times {\cal Q}_0
 + \frac{1}{N^2} \times {\cal Q}_2 \right)
\end{eqnarray*}
where the ${\cal G}_i$ and ${\cal Q}_i$ represent squared
sub-amplitudes for the 2-quark and 4-quark processes respectively.
Preliminary investigations suggest that the leading piece, ${\cal G}_0$
is both a good approximation to the full matrix element (including both
processes) and considerably faster to run. For this reason, the
inclusion of the term ${\cal G}_0$ has been the primary focus
of the $W+2$~jet calculation so far.

\section{Conclusions}
 
We have found large radiative corrections to the $Wb{\bar b}$
and $Zb{\bar b}$ processes. These results can significantly change
estimates of the backgrounds to the processes $p{\bar p} \to WH$
and  $p{\bar p} \to ZH$, which will be
important search channels at the Tevatron.
For the more general calculation of $W/Z +2$~jet production,
work is still ongoing but first results should be available soon.

\section*{Acknowledgements}

I would like to thank my collaborator Keith Ellis. An EU grant
to attend the Moriond conference is gratefully acknowledged.

\end{document}